\documentclass[10pt]{article}
\usepackage{times}
\usepackage{mathfont}

\newtheorem{lemma}{Lemma}

\newtheorem{theorem}{Theorem}

\newcommand{\qed}{~$\Box$\medbreak}
\newenvironment{proof}{\noindent{\bf Proof: }}{\qed}

\DeclareSymbolFont{lasy}{U}{lasy}{m}{n}
\let\Box\undefined
\DeclareMathSymbol\Box{0}{lasy}{"32}

\usepackage{url}
\input epsf
\def\efig#1#2{\hbox{\epsfxsize=#1\epsfbox{#2}}}

\makeatletter
\long\def\@makecaption#1#2{
   \vskip 10pt 
   \setbox\@tempboxa\hbox{{\small #1. #2}}
   \ifdim \wd\@tempboxa >\hsize   
       {\small #1. #2}\par        
     \else                        
       \hbox to\hsize{\hfil\box\@tempboxa\hfil}  
   \fi}

\def\@begintheorem#1#2{\it\trivlist
				\item[\hskip \labelsep{\bf #1\ #2.\ }]}
\def\@opargbegintheorem#1#2#3{\it\trivlist
				\item[\hskip \labelsep{\bf #1\ #2\ {\rm(#3)}.}]}

\makeatother

\begin{document}

\let\tilde~
\def~{$\sim$}

\title{Linear Complexity Hexahedral Mesh Generation}

\author{David Eppstein\thanks{Department of Information and
Computer Science, University of California, Irvine, CA 92697-3425, USA,
http://www.ics.uci.edu/~eppstein/,
eppstein@ics.uci.edu.
Work supported in part by NSF
grant CCR-9258355 and by matching funds from Xerox Corp.
A preliminary version of this paper appeared at the 12th ACM Symp. on
Computational Geometry.}}

\date{ }
\maketitle   

\let~\tilde

\begin{abstract}
We show that any polyhedron forming a topological ball
with an even number of quadrilateral sides
can be partitioned into $O(n)$ topological cubes, meeting face to face.
The result generalizes to non-simply-connected polyhedra
satisfying an additional bipartiteness condition.
The same techniques can also be used to reduce the geometric 
version of the hexahedral mesh generation problem to a finite case analysis
amenable to machine solution.
\end{abstract}

\section{Introduction}

There has recently been a great deal of theoretical work on unstructured
mesh generation for finite element methods, largely concentrating on
triangulations and higher dimensional simplicial complexes;
see \cite{BerEpp-CEG-95} for a survey of these results.
However in the numerical community, where these meshes have been
actually used, meshes of quadrilaterals or hexahedra (cuboids) are often
preferred due to their numerical properties~\cite{BenPerMer-IMR-95}.  For this
reason many mesh generation researchers are working on systems for
construction of hexahedral meshes.  There has also been some theoretical
work on quadrilateral and hexahedral meshes~\cite{BenBlaMit-SCG-95,
BerEpp-IMR-97, Mit-WCG-95, MulWei-ESA-97, RamRamTou-CCCG-95} but much more
remains to be done.

\begin{figure*}[tp]
$$\efig{3in}{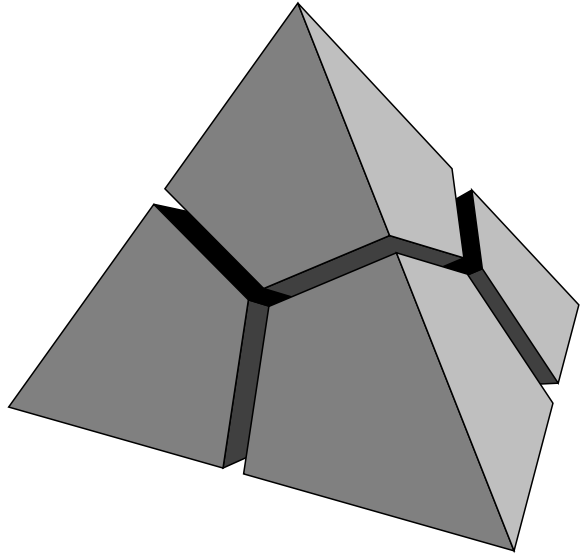}$$
\caption{Tetrahedron partitioned into four hexahedra.}
\label{hexedtet}
\end{figure*}

\begin{figure*}[tp]
$$\efig{2in}{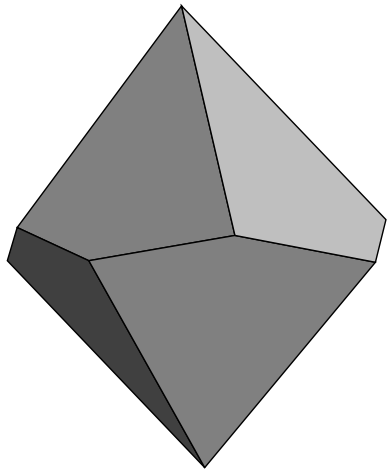}$$
\caption{Can this octahedron be meshed with convex hexahedra?}
\label{octahedron}
\end{figure*}

There is a straightforward method for generating hexahedral meshes, if
one allows the mesh to include additional {\em Steiner points} as
vertices.  Simply find a Steiner tetrahedralization of the
domain~\cite{ChaPal-DCG-90}, then subdivide each tetrahedron into four
hexahedra as shown in Figure~\ref{hexedtet}.  But as
Mitchell~\cite{Mit-WCG-95} notes, this type of boundary subdivision can
make it difficult to mesh several adjoining domains simultanously
(such internal domain boundaries can arise either because the problem is
defined in terms of multiple domains, or as part of a parallel mesh
generation process).  Further, splitting tetrahedra produces hexahedra
with some very sharp angles and some very obtuse angles, which are hard to
improve in a later mesh smoothing stage, especially when they occur on the
domain boundary.

We consider here a common variant of the hexahedral mesh generation
problem, in which we avoid some of these problems by restricting the
location of new Steiner points to the {\em interior} of the domain.  The
boundary (which is assumed to be a planar quadrilateral mesh) must remain
unsubdivided.  Although various authors have studied heuristics for this
version of hexahedral mesh generation, its theoretical properties are
not well understood and pose many interesting problems.  In particular
the computatational complexity of determining whether a polyhedron
admits a mesh of convex hexahedra respecting the polyhedron's boundary
is unknown.  Even some very simple cases, such as the eight-sided
polyhedron shown in Figure~\ref{octahedron}, remain open~\cite{Sch-Open};
the hexahedral meshes known for this octahedron are very complicated
and involve nonconvex or degenerate cuboids.

For the planar case, the corresponding problem is easy:
a polygon can be subdivided into convex quadrilaterals, meeting edge to
edge, without extra subdivision points on the boundary, if and only if
the polygon has an even number of sides.
One can efficiently find a set of $O(n)$ Steiner points that suffice
for this problem \cite{RamRamTou-CCCG-95},
and there has been some progress on finding the minimum possible number
of Steiner points for a given polygon \cite{MulWei-SCG-97}.

Thurston~\cite{Thu-93} and Mitchell~\cite{Mit-WCG-95} independently
showed a similar characterization for the existence of hexahedral meshes,
with some caveats.  First, the polyhedron to be meshed has to be a
topological ball (although the method generalizes to certain polyhedra
with holes).  And second, the mesh is {\em topological}: the elements
have curved boundaries and are not necessarily convex. However they must
still be combinatorially equivalent to cubes, and must still meet face to
face (i.e. any internal boundary between elements must be a face of
both elements; we formalize this requirement in the next section by
requiring the elements to form a cell complex).  Thurston and Mitchell
both showed that any polyhedron forming a topological ball has a
topological hexahedral mesh, without further boundary subdivision, if and
only if there are an even number of boundary faces all of which are
quadrilaterals. (Indeed, even parity of the number of faces is a
necessary condition for the existance of cubical meshes in any dimension,
regardless of the connectivity of the input, since each individual cube
has evenly many faces which either contribute to the boundary or are
paired up in the interior.) However Thurston and Mitchell's method may
produce meshes with high complexity (nonlinear in the number of features
of the original polyhedral domain).

In this paper we discuss an alternate method for hexahedral grid
generation.  Our method combines refinement of a tetrahedral mesh with
some local manipulation near the boundary based on planar graph theory.
It is similar in spirit to mesh generation heuristics of
Schneiders~\cite{SchBun-CAGD-95} and others, in which one first fills the
interior of the domain with cubes before attempting to patch the remaining
regions between these cubes and the boundary.  A similar idea of
``buffering'' the boundary of the input from subdivisions occurring in
the interior was also used in a tetrahedralization algorithm of
Bern~\cite{Ber-SCG-93}.

Our new hexahedral meshing technique has three advantages over that of
Mitchell and Thurston.
First, we prove an $O(n)$ bound on the number of cells needed for a
topological hexahedral mesh.  Second, because our method avoids duality,
it seems easier to extend it to the more practically relevant geometric
version of the mesh generation problem, in which the elements must be
convex cuboids: we exhibit a finite collection of polyhedra (formed by
subdividing the boundary of a cuboid) such that if these polyhedra can all
be geometrically meshed, any polyhedron forming a topological ball with an
even number of quadrilateral sides can also be geometrically meshed, with
$O(n^2)$ cells.  Third, the method generalizes to a different class of
non-simply connected polyhedra than those handled by Mitchell and
Thurston's method.

\begin{figure*}[tp]
$$\efig{2.5in}{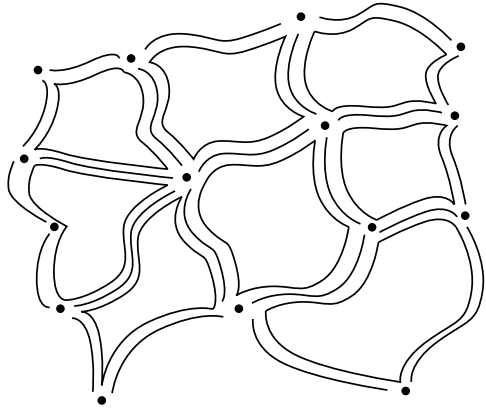}$$
\caption{Topological quadrilateral mesh, showing individual
faces, edges, and vertices.}
\label{ccx}
\end{figure*}

These results are not practical in themselves: the number of elements is
too high and we have not satisfactorily completed the solution to the
geometric case.  Practical hexahedral mesh generation methods are still
largely heuristic and will often fail or require the input boundary to
be modified.  There is a possibility here of a two way interaction
between theory and practice: as heuristic mesh generators improve they
may soon be good enough to solve the finite number of cases remaining in
our geometric mesh generation method, and thereby prove that all
even-quadrilateral polyhedra can be meshed.  In the other direction,
even an impractical proof of the existence of meshes can be helpful, by
guaranteeing that an incremental heuristic method such as the
whisker-weaving idea of Benzley et al.~\cite{BenBlaMit-SCG-95} will not
get stuck in a bad configuration.

\section{Statement of the problem}

Let us define more formally the topological mesh generation problem
solved here.  We assume we are given a {\em domain} topologically
equivalent to the closed ball in three dimensions (later we will
consider other more complicated domain topologies).  The boundary of the
domain is assumed to be covered by a finite {\em cell complex}; that is,
a collection of finitely many {\em cells}: sets equivalent to closed
balls of various dimensions, with disjoint relative interiors, such that
the boundary of any cell is covered by lower dimensional cells,
and any nontrivial intersection of two cells is itself a cell.
Our task is to extend this partition to a finite cell complex covering
the overall domain.

We assume the boundary cell complex is a {\em quadrilateralization}; in
particular, every {\em edge} (one-dimensional cell) has two {\em
vertices} (zero-dimensional cells) as its endpoints, and every {\em
face} (two-dimen\-sional cell) has a cycle of four distinct edges as its
boundary. (Figure~\ref{ccx}.)

We wish the cell complex produced by our algorithm to be a {\em
hexahedral mesh}: the same conditions as above apply to every edge and
face of the complex, but in addition every three-dimensional cell must
be a {\em hexahedron}: it should have six quadrilateral faces on its
boundary, meeting in edges and vertices with the same combinatorial
structure as the faces, edges, and vertices of a cube.

Any simply-connected polyhedron with quadrilateral faces satisfies the
input conditions.  Any partition of that polyhedron into convex cuboids
meeting face-to-face satisfies our output conditions.  However our
definition also allows partitions into non-convex and non-polyhedral
cells.

\section{Thurston and Mitchell}

\begin{figure*}[tp]
$$\efig{3in}{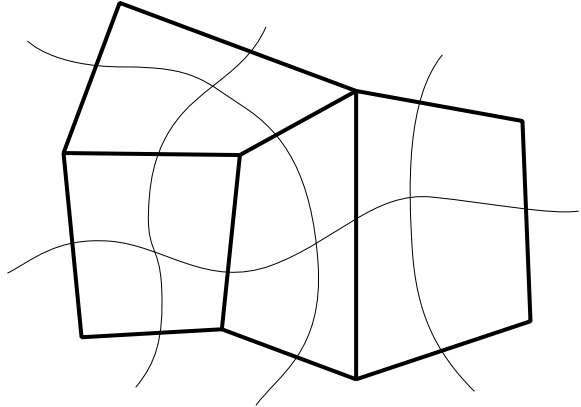}$$
\caption{Quadrilateral mesh and dual curves.}
\label{quaddual}
\end{figure*}

\begin{figure*}[tp]
$$\efig{2.5in}{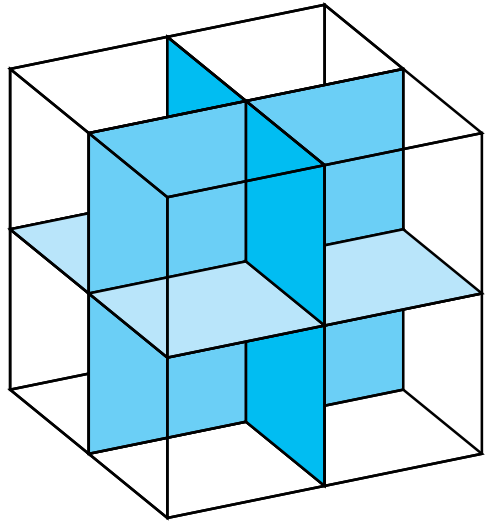}$$
\caption{Hexahedron and dual surfaces.}
\label{hexdual}
\end{figure*}

Before describing our own methods, we briefly discuss
those of Thurston~\cite{Thu-93} and Mitchell~\cite{Mit-WCG-95}.

The method of both these authors is to treat a hexahedral mesh
as being the dual
to an arrangement of surfaces~\cite{BenBlaMit-SCG-95}, and a quadrilateral
mesh such as the one on the boundary of the polyhedron
as being the dual to an arrangement of curves.
Given a mesh of quadrilaterals, one can find this curve arrangement
simply by connecting the midpoints of opposite sides of each
quadrilateral by curves (Figure~\ref{quaddual}).  Similarly, given a
mesh of hexahedra,
one can find these dual curves on each boundary facet of each hexahedron,
and connect them by quadrilaterals to form surfaces, meeting in triple
points at the center of each hexahedron (Figure~\ref{hexdual}).

The problem then becomes one of performing the opposite transformation:
extending the given surface curve
arrangement to an interior surface arrangement, and then finding a
collection of topological hexahedra dual to this surface arrangement.

Thurston and Mitchell solve the first part of this problem (extending
the boundary curve arrangement to an interior surface arrangement)
by extending curves with an even number of self-intersections
to surfaces independently of each other; they pair up curves
with an odd number of self-intersections and form a surface for each pair.

For the second part (transforming these surfaces to dual hexahedra),
note that, for surfaces in general position,
the structure of the surface arrangement can be represented
as a collection of vertices (for each triple intersection of surfaces)
and edges (segments of pairwise intersections of surfaces).
Each vertex should correspond to a dual hexahedron and each
edge should correspond to a pair of hexahedra sharing a common face.
However not every set of surfaces determines a dual cell complex in this way,
because some of the vertices may be connected by multiple edges,
resulting in hexahedra that do not intersect in a single face
and violating our requirement that the intersection of two cells be
another cell. Thurston and Mitchell solve this problem by
surrounding problematic regions of the surface arrangement with spheres
in a way that causes these multiple adjacencies to be removed.
The result is a collection of surfaces with a dual hexahedral mesh
solving our mesh generation problem.

\begin{theorem}[Mitchell, Thurston]
Any simply connected three-dimensional domain with an even number of
quadrilateral boundary faces can be partitioned into a hexahedral mesh
respecting the boundary.
\end{theorem}

\begin{figure*}[t]
   \centering
   \begin{tabular}{cc}
      \efig{2.5in}{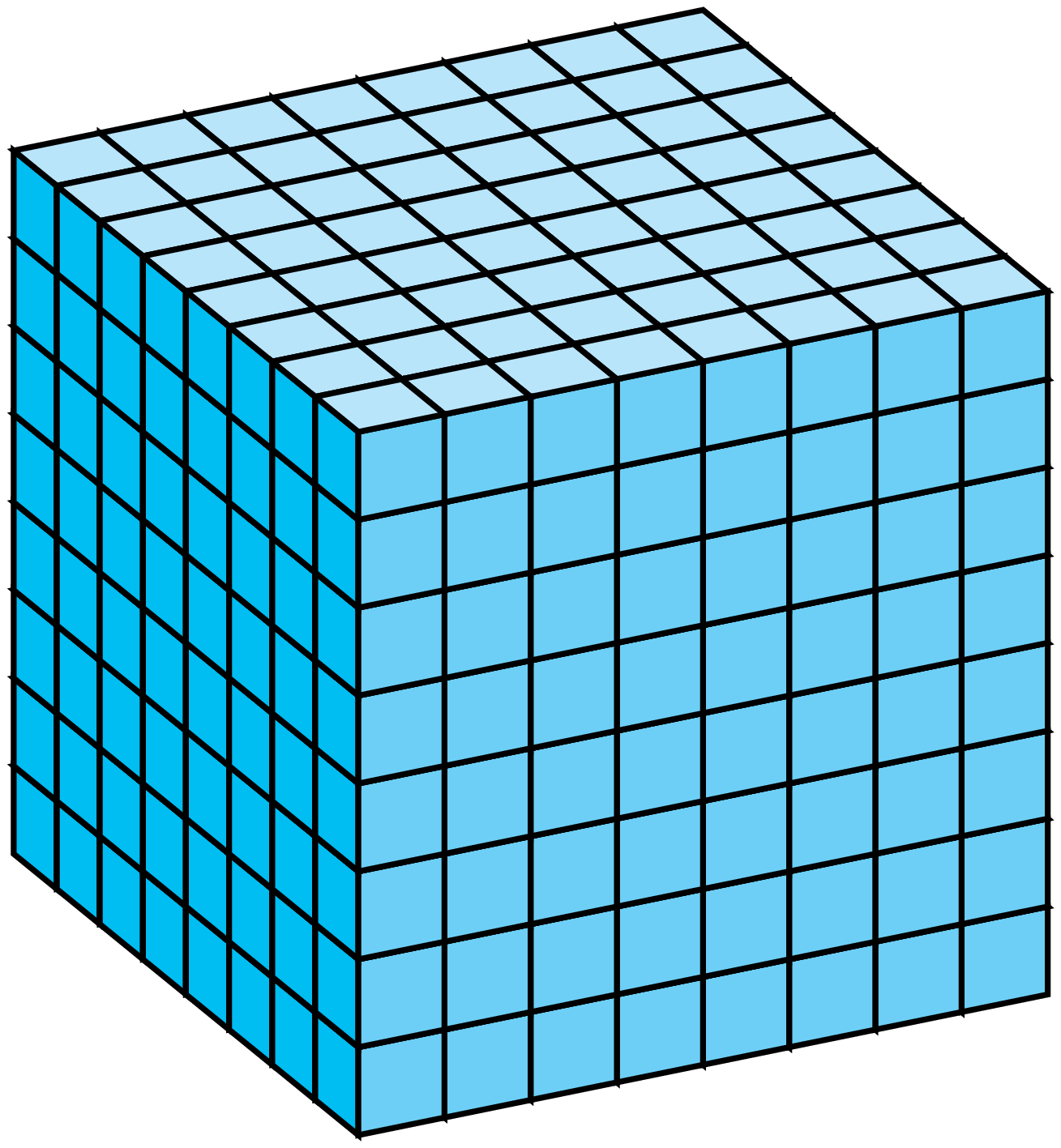}&
      \efig{2.5in}{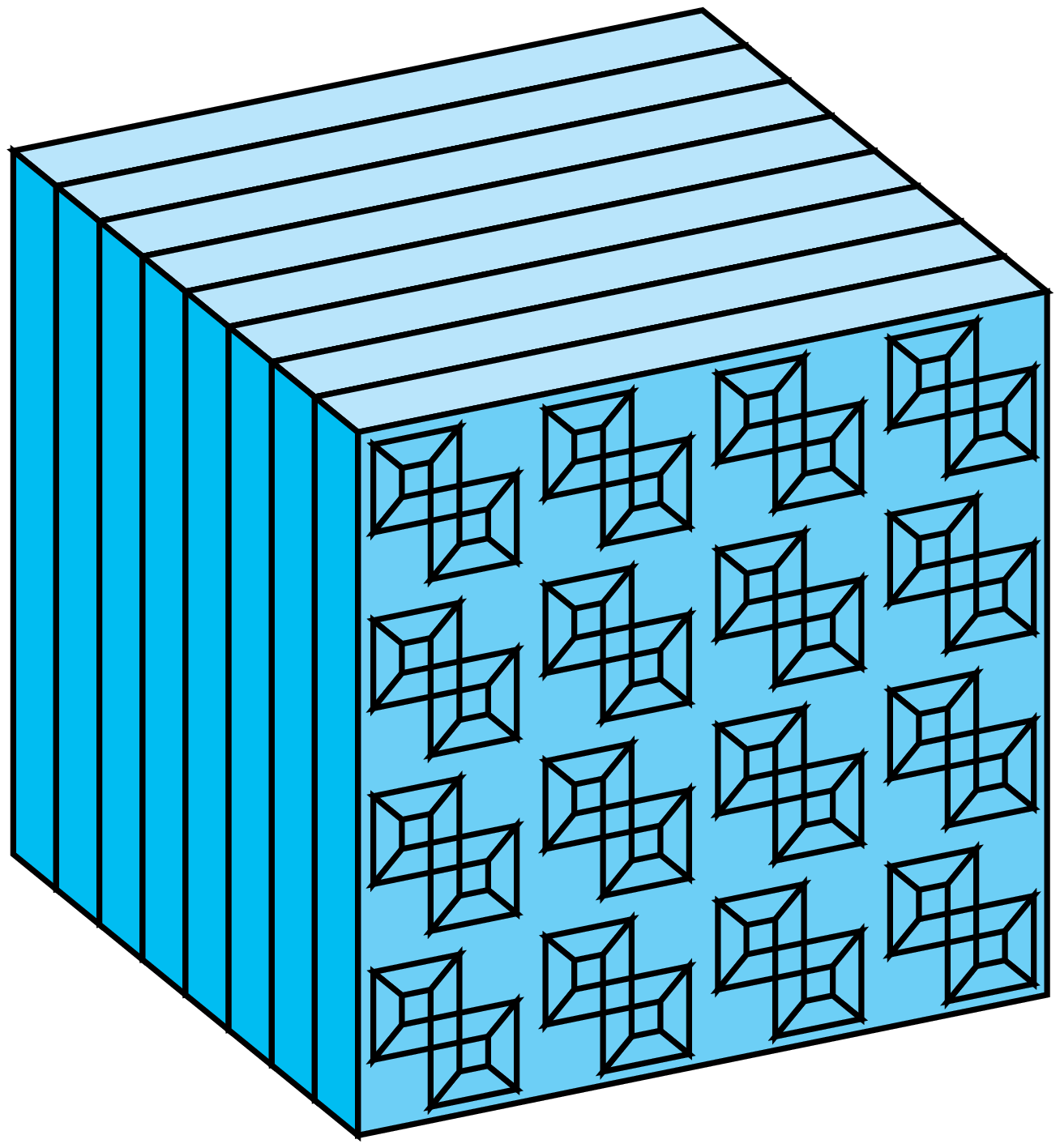}\\
   \end{tabular}
   \caption{Bad examples for Thurston and Mitchell:
	    $\Omega(n^{3/2})$ (left) and
	    $\Omega(n^2)$ (right).}
   \label{ThuMitch-lb}
\end{figure*}

However this method does not provide much of a guarantee on the
{\em complexity} of the resulting mesh, that is, of the number
of hexahedral cells in it.  This complexity is very important, as it
directly affects the time spent by any numerical method using the mesh;
even small constant factors can be critical.

It is not hard to provide examples in which this dual surface
method constructs meshes with more than linearly many elements
(measured in terms of the complexity of the polyhedron boundary).
We provide two (Figure~\ref{ThuMitch-lb}):
\begin{itemize}
\item A cube in which each square is subdivided into an
$O(\sqrt n)$ by $O(\sqrt n)$ grid.
The boundary curve arrangement dual to these grids consists of
$\Omega(\sqrt n)$ Jordan curves without self-intersections.
The method of Thurston and Mitchell extends each of these curves
to a plane; these $\Omega(\sqrt n)$ planes will have
$\Omega(n^{3/2})$ triple intersection points, so this method
will produce a mesh of total
complexity $\Omega(n^{3/2})$.
\item
A cube in which four of the boundary squares are subdivided into
$\Omega(n)$ rectangles, forming $\Omega(n)$ disjoint
and non-self-intersecting Jordan curves, and in which the remaining two
squares have $\Omega(n)$ quadrilaterals arranged in a pattern to form
$\Omega(n)$ disjoint curves with one self-intersection each.
The Jordan curves will be extended to planes;
if one incautiously matches the one-inter\-sec\-tion curves
into pairs, with each pair consisting of one curve from each side of the
cube, the surfaces formed by extending these pairs will each cross
all $\Omega(n)$ of the planes coming from the Jordan curves, and form a
mesh with overall complexity $\Omega(n^2)$.
\end{itemize}

\section{Linear-complexity mesh generation}

As we saw above, the mesh generation method of Thurston and Mitchell
can produce meshes with $\Omega(n^{3/2})$ or $\Omega(n^2)$ hexahedra.
We now describe our new topological mesh generation method, which will always
give meshes with $O(n)$ complexity.

\begin{figure*}[tp]
$$\efig{2.5in}{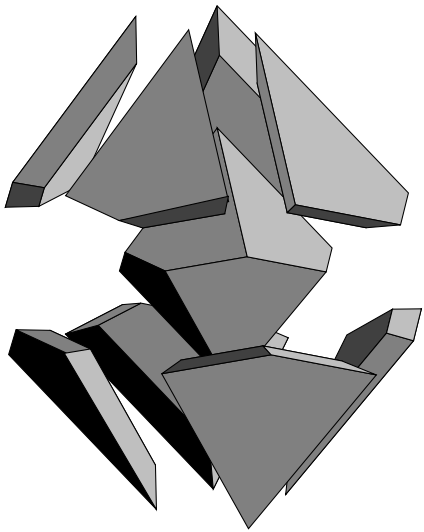}$$
\caption{Separation of boundary from interior by buffer layer.}
\label{tiled-oct}
\end{figure*}

\begin{figure*}[tp]
$$\efig{2.5in}{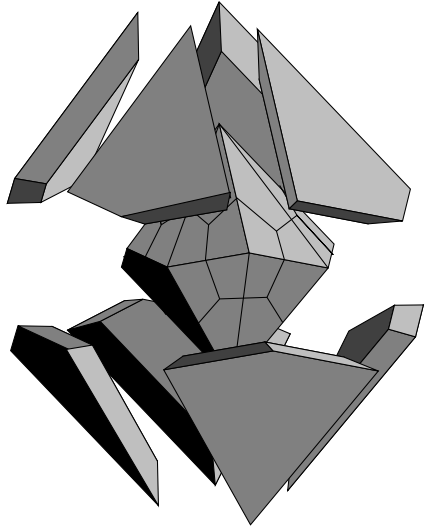}$$
\caption{Hexahedralization of interior.}
\label{hexed-oct}
\end{figure*}

\begin{figure*}[tp]
   \centering
   \begin{tabular}{cc}
      \efig{1.5in}{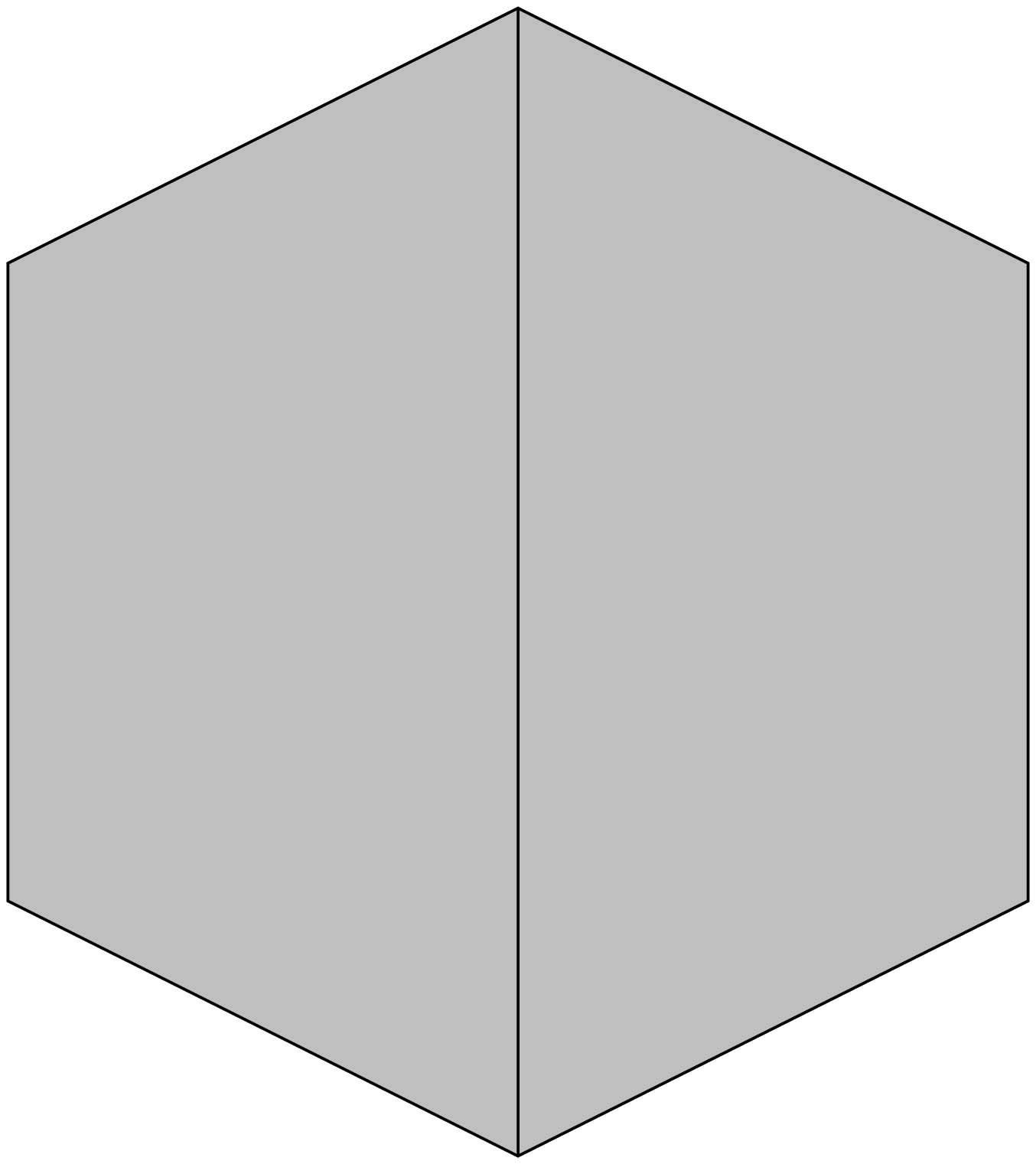}&
      \efig{1.5in}{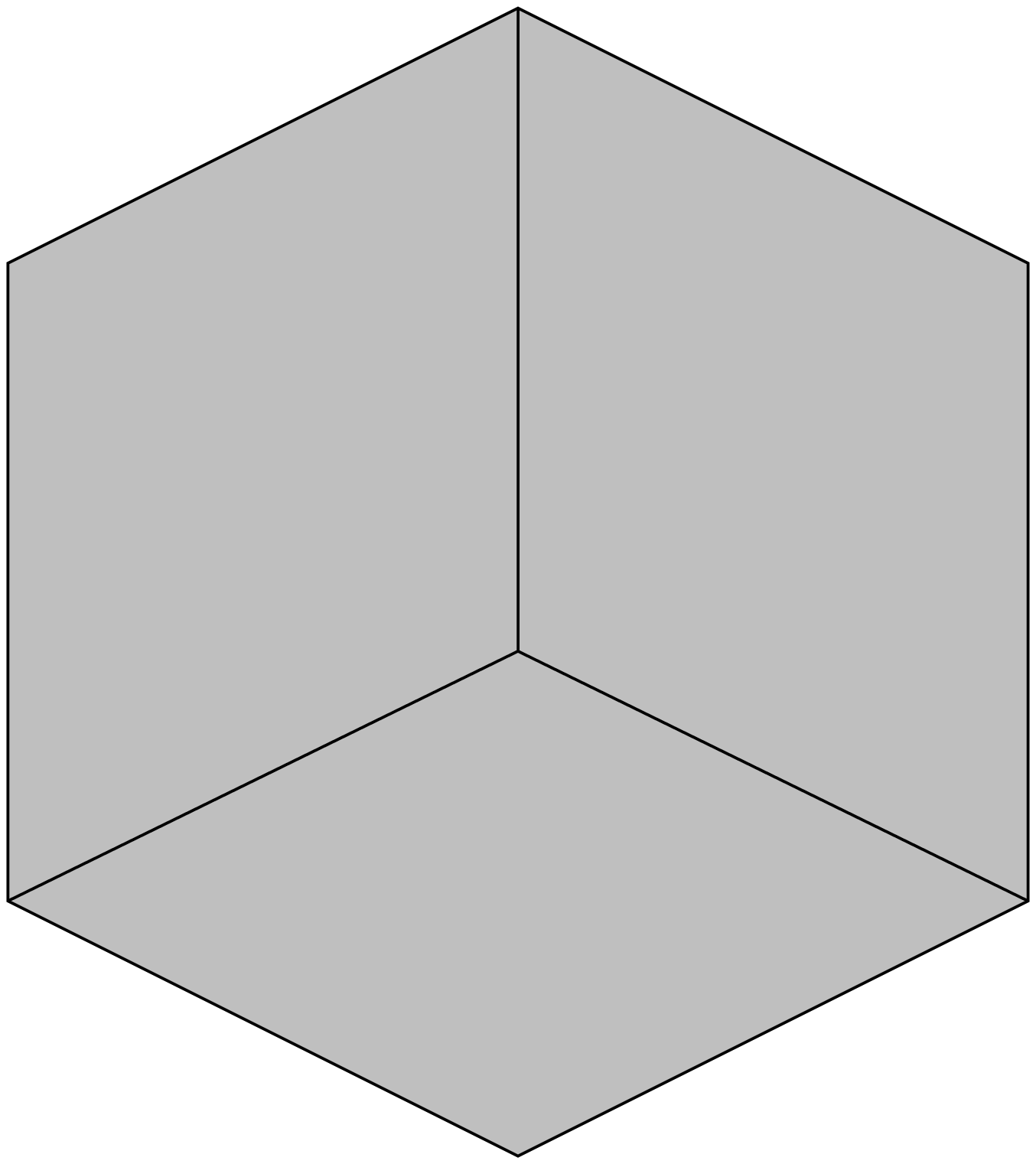}\\
   \end{tabular}
   \caption{Partition of hexagons into two and three quadrilaterals.}
   \label{hexquad}
\end{figure*}

Our method has the following main steps:
\begin{enumerate}
\item We separate the boundary $B$ of the polyhedron from its interior by a
``buffer layer'' of cubes.  We do this by finding a surface $S$ inside
the polyhedron, isomorphic to the polyhedron's boundary, and sitting in
the same orientation.  We then connect corresponding pairs of vertices
on the two surfaces with edges, corresponding pairs of edges on the two
surfaces with quadrilateral faces spanning pairs of connecting edges,
and corresponding pairs of faces on the two surfaces with hexahedra
(Figure~\ref{tiled-oct}).

\item We triangulate the inner surface of the buffer layer, and
tetrahedralize the region inside this triangulated surface.
A (topological) tetrahedralization with $O(n)$ complexity can be found
by connecting each triangle on $S$ to a common interior vertex.

\item We split each interior tetrahedron into four hexahedra
as in Figure~\ref{hexedtet}. This subdivision should be
done in such a way that any two tetrahedra that meet in a facet or edge
are subdivided consistently with each other. As a result, each edge and
face in $S$ becomes subdivided, and each quadrilateral connecting $B$ to
$S$ gains an additional subdivision point and so becomes combinatorially
equivalent to a pentagon (Figure~\ref{hexed-oct}).

\item Because $B$ is by assumption a planar graph with all faces even,
it is bipartite.  (The well-known fact that even faces implies
bipartiteness is the planar dual to the fact that even vertex degree
implies the existence of an Euler tour, but it can easily be proved
directly.)  Let $U$ and $V$ be the two vertex sets of a bipartition of
$B$ (without loss of generality, $|U|<|V|$).  Each vertex of $B$
corresponds to an edge in the buffer layer connecting $B$ to $S$.  We
subdivide the subset of those edges corresponding to vertices in $U$.
Each of the quadrilaterals
connecting $B$ to $S$ has one such edge, so (together with the subdivided
edge in $S$ itself) these subdivisions cause each such quadrilateral to
become combinatorially a hexagon.

\item After these subdivision steps, each of the cells in the buffer
layer is now combinatorially
a polyhedron with seven quadrilateral facets and four hexagon facets.
We subdivide the hexagons into either two or three quadrilaterals each,
as shown in Figure~\ref{hexquad}.
We explain below how to do this in such a way that each cell of the
buffer layer has an odd number of hexagons subdivided into each type;
either one hexagon is subdivided into two quadrilaterals and three hexagons
are subdivided into three quadrilaterals each, or
three hexagons are subdivided into two quadrilaterals and one hexagon
is subdivided into three quadrilaterals.

\item At this point, all the buffer cells are combinatorially polyhedra
with either 16 or 18 quadrilateral boundary facets.  We partition each
cell into a mesh of $O(1)$ hexahedra.  (The existence of such a mesh is
guaranteed by Mitchell and Thurston's results.  Alternately, if the
triangulation of $S$ is chosen carefully using the same bipartition used
above, there will only be two combinatorial types of cell; it is an
amusing exercise to fill out these cases by hand.)
\end{enumerate}

The remaining step that has not been described is how we choose
whether to subdivide each of the faces connecting $B$ to $S$ into two or
three quadrilaterals, so that each buffer cell has an odd number of
subdivided faces of each type.

\begin{figure*}[t]
   \centering
   \begin{tabular}{cc}
      \efig{3in}{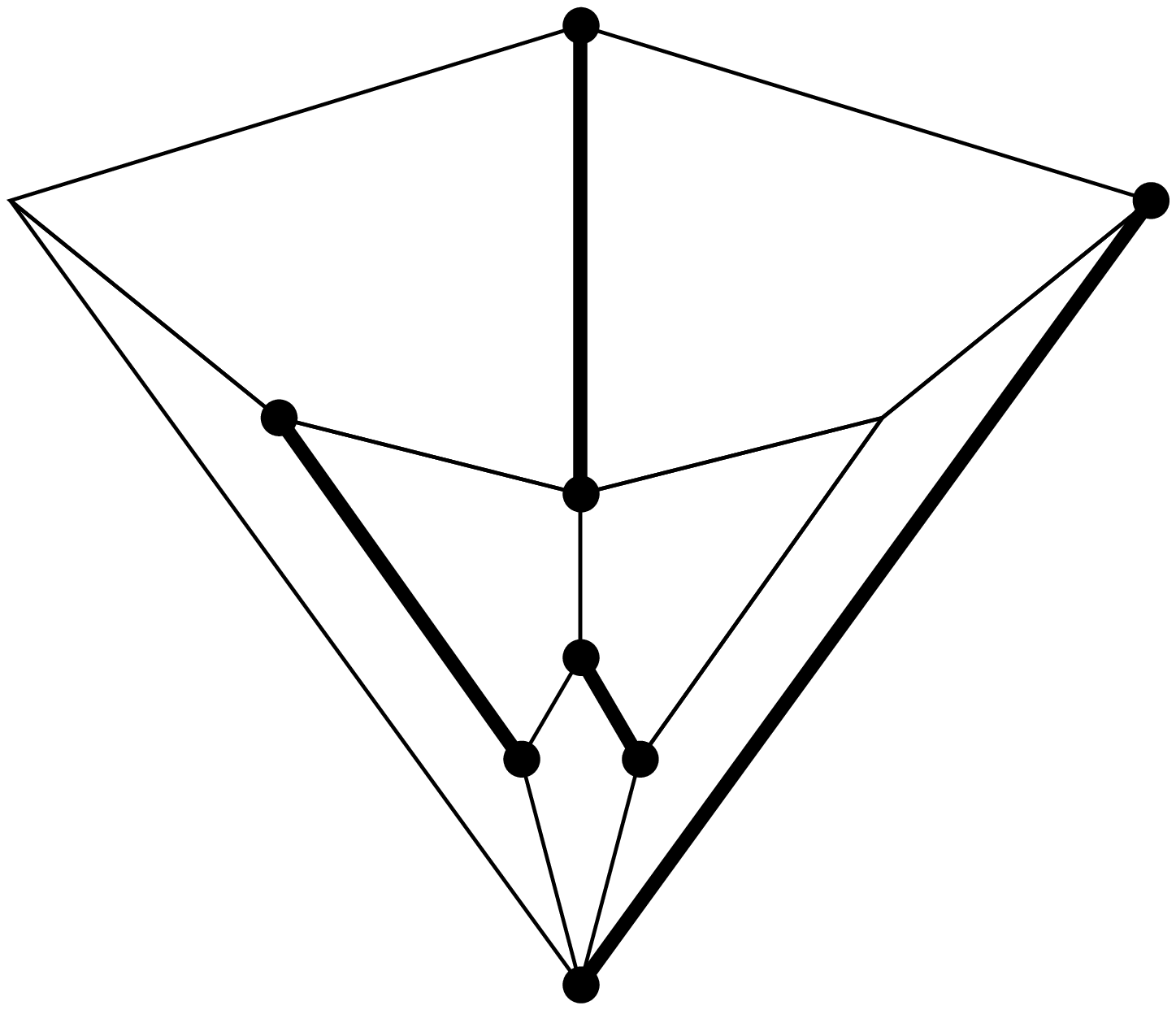}&
      \efig{2in}{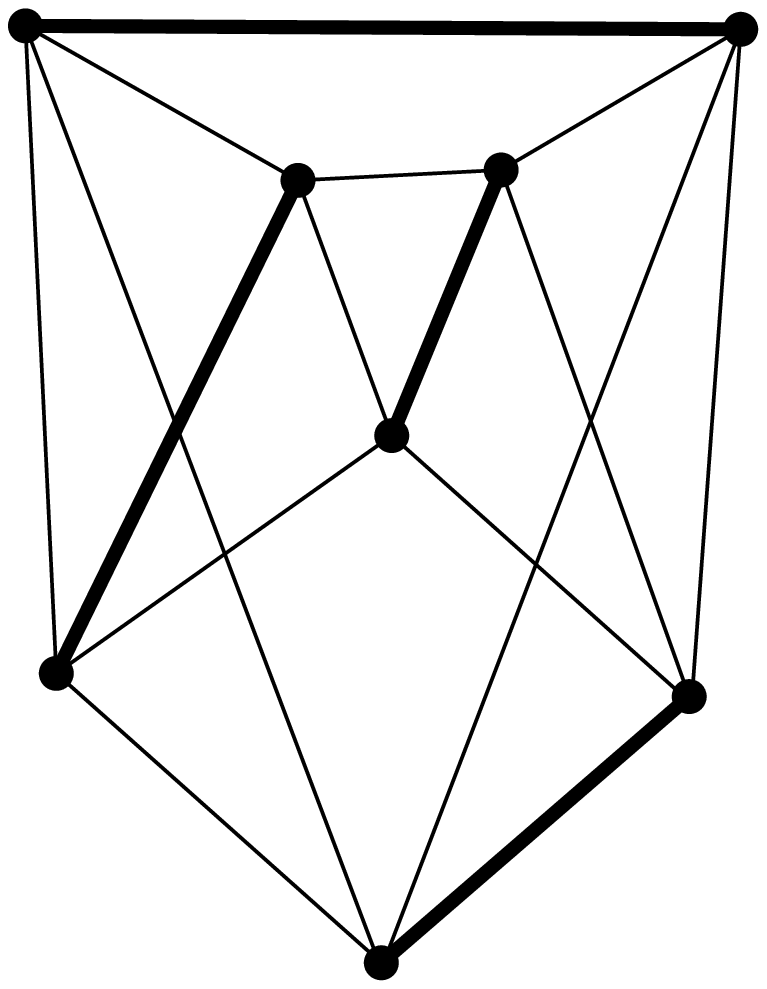}\\
   \end{tabular}
   \caption{Quadrilateral mesh (left) and its dual (right).
The bottom vertex of the dual corresponds to the outer face of the mesh;
other dual vertices are arranged roughly in the same positions as the
mesh faces. The highlighted edges form a minimum-weight matching in the
dual, and  a corresponding set of edges in the mesh meeting each
quadrilateral an odd number of times.}
   \label{chinese}
\end{figure*}

\begin{lemma}
Given any planar graph $G$ with evenly many quadrilateral faces,
we can in polynomial time find a set $S$ of edges of $G$
such that each face of $G$ is bounded by an odd number of edges in $S$.
\end{lemma}

\begin{proof}
We use a technique familiar from the solution to the Chinese
postman problem.
Construct the dual graph $G'$, and form a metric
on the vertices of $G'$ with distances equal to the
lengths of shortest paths in $G'$.  By assumption $G'$ has an even number
of faces, so $G'$ has evenly many vertices and there are perfect
matchings in this metric; take the
minimum weight perfect matching.  This corresponds to a collection
of paths in $G'$; any two paths must be edge-disjoint
since otherwise one could perform a swap and find a shorter matching.
The union of these paths is a subgraph $S'$ of $G'$
(actually a forest) in which every vertex has odd degree.
(Each vertex has odd degree in the path in which it is an endpoint, and
even
degree in all other paths, so the total is odd.)
Taking the corresponding edges in $G$ gives a set $S$ of edges
having an odd number of incidences with each face, as was required in
the statement of the lemma.
\end{proof}

In fact, it is not hard to see that
this procedure finds the set $S$ of smallest cardinality.
This process is depicted in Figure~\ref{chinese},
which depicts for the skeleton of the polyhedron in
Figure~\ref{octahedron} the minimum weight matching on the shortest-path
metric of the dual graph, and the resulting set of edges on the original
graph. In this example, the paths in $G'$ coming from the matching consist
of a single edge each.

We apply this lemma to choose how to subdivide the hexagonal faces of
the buffer layer between $B$ and $S$.
Recall that $B$ is a planar graph with evenly many quadrilateral faces.
Further, each face of $B$ corresponds to a cell of the buffer layer,
and edge of $B$ corresponds to one of the hexagons that we
wish to subdivide.
We use the method of the lemma to find a set $S$
of edges of $B$ incident an odd number of times to each face of $G$;
equivalently this corresponds to a set of hexagonal
faces incident an odd number of times to each cell in the buffer layer.
We subdivide that set of faces into three quadrilaterals each,
and the remaining faces into two quadrilaterals, as shown
in Figure~\ref{hexquad}.

We summarize the results of this section.

\begin{theorem}
Given any polyhedron $P$ forming a topological ball with an even number
$n$ of faces,
all quadrilaterals, it is possible to partition $P$ into $O(n)$ topological
cubes meeting face-to-face, such that each face of $P$ is a face of some cube.
\end{theorem}

\begin{proof}
The correctness of this method is sketched above.
If $P$ has $n$ faces, there are $2n$ tetrahedra in the interior of $S$,
subdivided into $8n$ hexahedra.  In addition the $n$ cubes connecting
$B$ with $S$ are subdivided into $O(1)$ hexahedra each, so the total
complexity is $O(n)$.
\end{proof}

\section{Geometric mesh generation}

We would like to extend the topological mesh generation method described
above to the more practically relevant problem of geometric mesh generation
(partition into convex polyhedra combinatorially equivalent to cubes).
Although our extension seems unlikely to be practical itself,
because of its high complexity and the poor shape of the hexahedra it produces,
it would be of great interest to complete a proof that all
polyhedra (with evenly many quadrilateral faces) can be meshed.
Also, it might make sense to include a powerful but impractical
theoretical method as part of a more heuristic mesher, to deal with
the difficult cases that might sometimes arise.

In any case, we have made some progress towards a geometrical
mesh generation algorithm, but have not solved the entire problem.
We have been able to solve the seemingly harder unbounded parts
of the problem, leaving only a bounded amount of case analysis to be done.
It seems likely that heuristic mesh generation methods may soon
be capable of performing this case analysis and finishing the proof.

We go through the steps of our topological mesh generation algorithm,
and describe for each step what changes need to be made to
perform the analogous step in a geometric setting.
However since our results here are incomplete, we do not fill in the
method in too much detail.

\begin{figure*}[tp]
   \centering
   \begin{tabular}{c}
      \efig{4in}{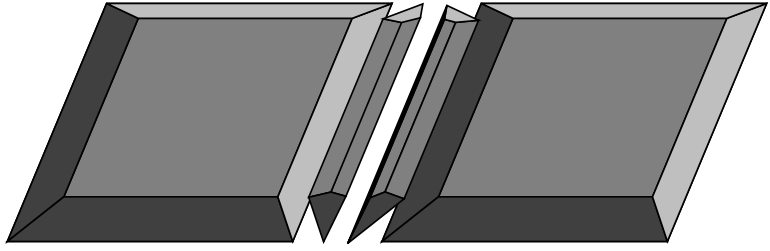}\\
   \end{tabular}
   \caption{Geometric hexahedralization: boundary faces are covered by
flat ``bevelled'' hexahedra, with two more hexahedra covering each edge.}
   \label{fillet}
\end{figure*}

\begin{figure*}[tp]
   \centering
   \begin{tabular}{ccc}
      \efig{1.4in}{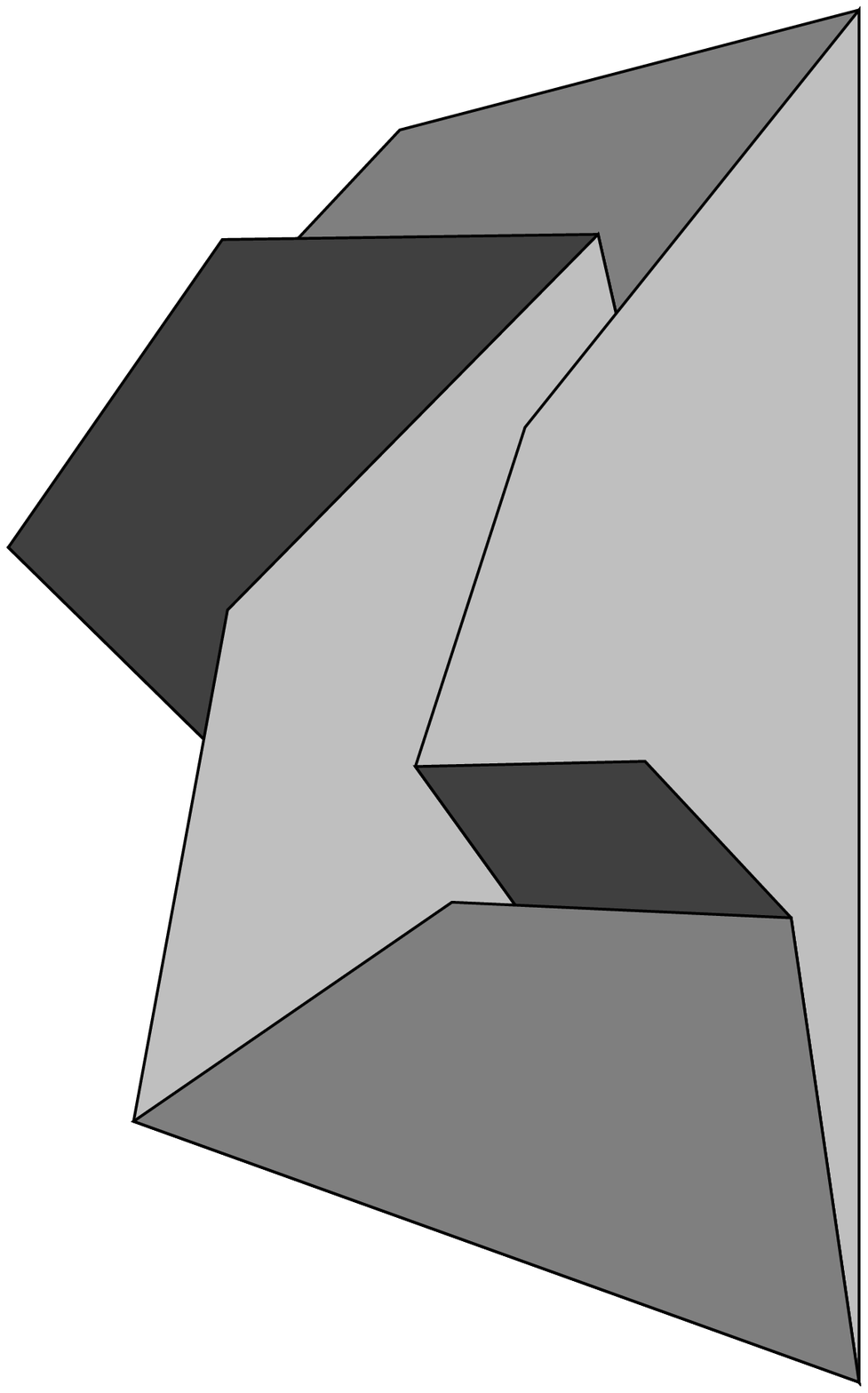}\qquad\qquad &
      \efig{1.4in}{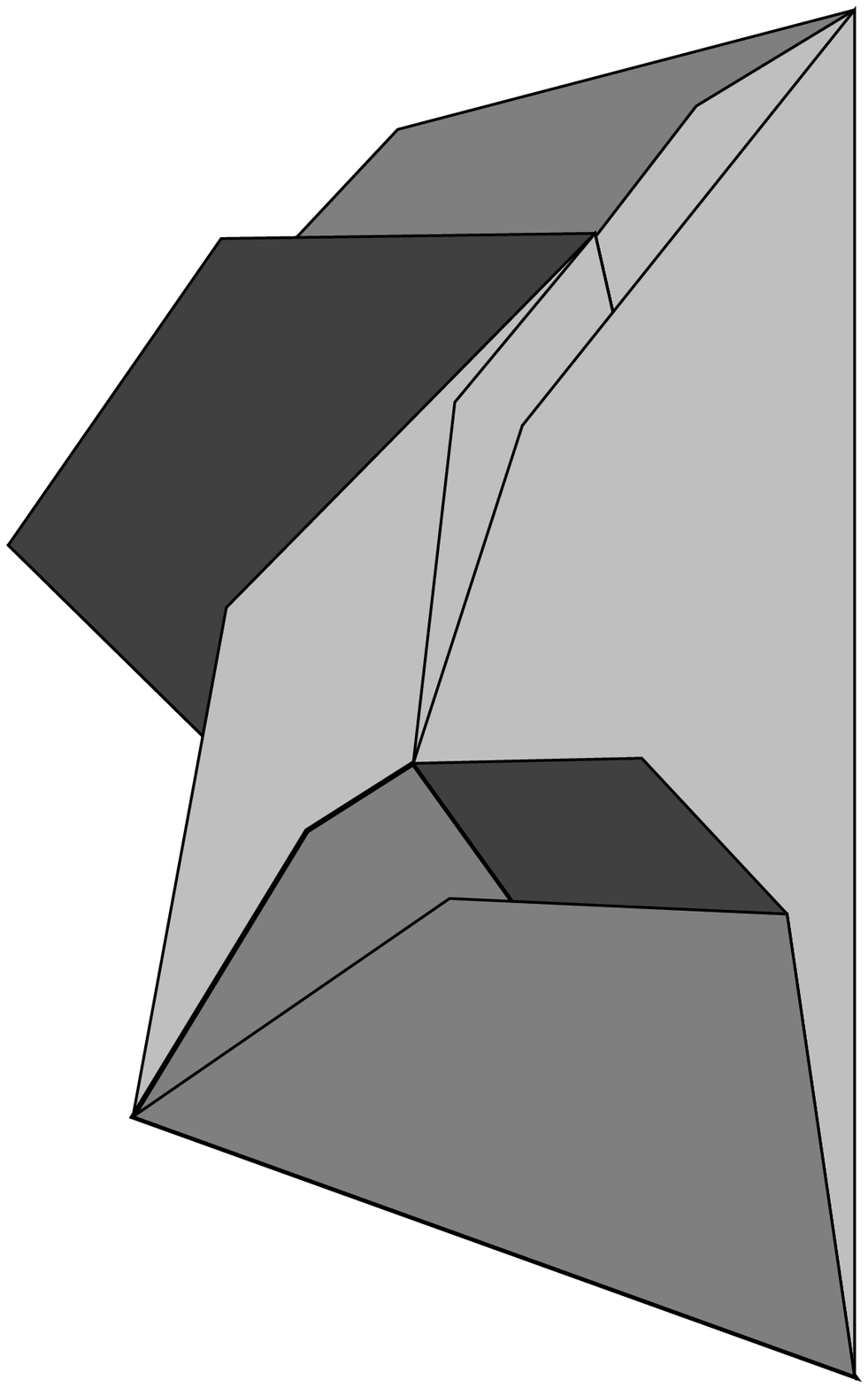}\qquad\qquad &
      \efig{1.4in}{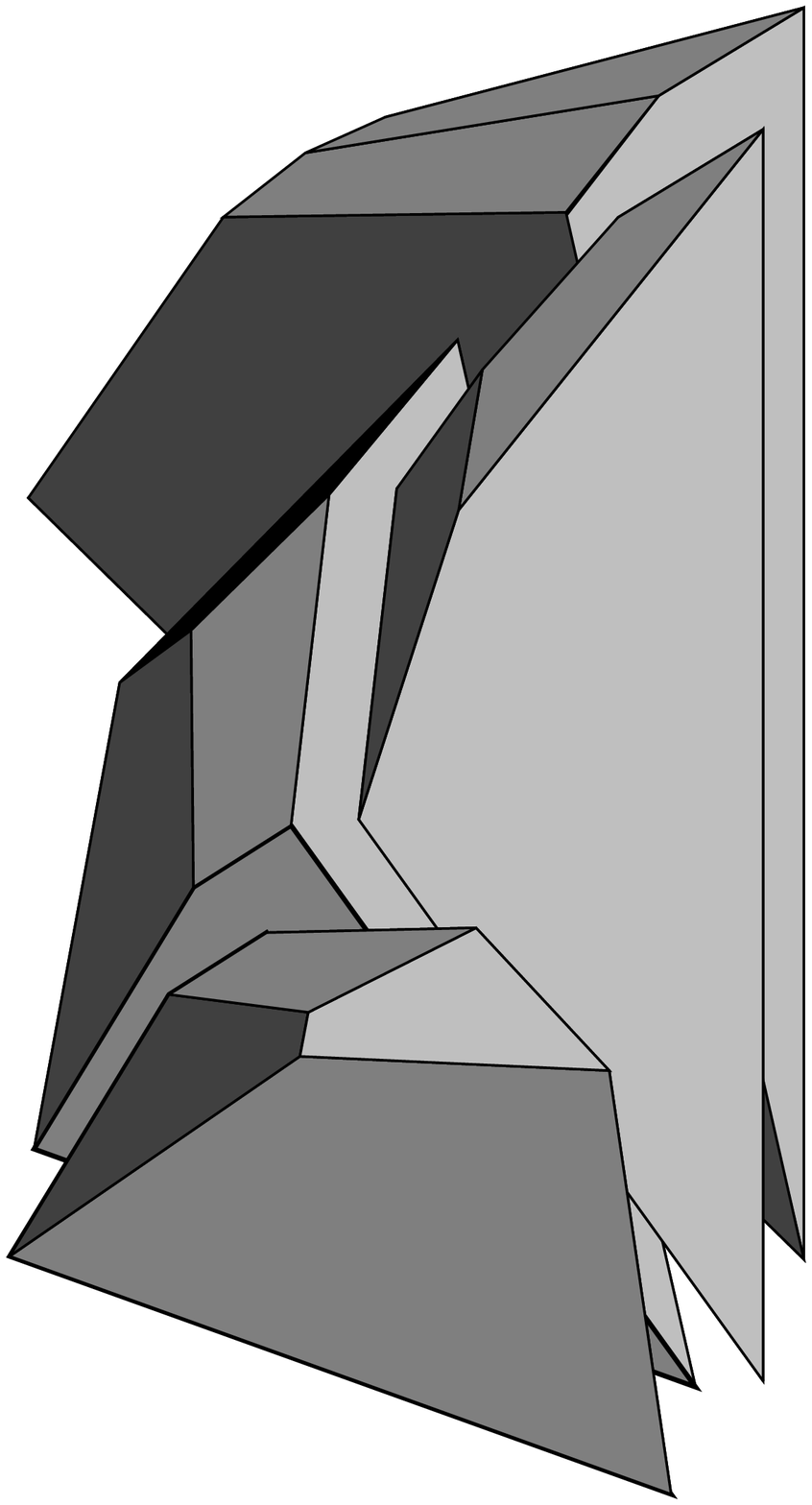}\\
   \end{tabular}
   \caption{Geometric hexahedralization: vertex figure is a polygon
bounded by kite-shaped quadrilaterals (left); adding more kites
triangulates the polygon (center); we form hexahedra from each
triangular cell (right).}
   \label{kites}
\end{figure*}

\begin{enumerate}
\item Our topological method separates the boundary $B$ of the
polyhedron from its interior by a single buffer layer of cuboids connecting
$B$ to an isomorphic surface $S$ inside the polyhedron.
Unfortunately there exist polyhedra 
for which no isomorphic interior surface can be connected to the boundary
by flat faces and straight edges; Figure~\ref{kites} shows an example
of a vertex surrounded by six quadrilaterals in such a way that, no matter
where the corresponding interior vertex is placed, some faces are
invisible to it and hence can not be connected by geometric hexahedra.
This example is easily completed to a polyhedron with the same property.
Instead we form a more complicated buffer layer in the following way.

We first cover each face $f$ of $B$ by a convex hexahedron, with the
opposite hexahedron face
parallel to $f$, very close to $f$, and somewhat smaller than $f$, so
that the other
four sides of $f$ are ``beveled'' to be nearly parallel to $f$.
As shown in Figure~\ref{fillet}, this causes the face to be replaced
by a set of five quadrilaterals, slightly indented into the polyhedron.

For any two faces $f$ and $f'$ sharing an edge of $B$, we then add to
our buffer layer two more cuboids, both also sharing the same edge,
connecting the sides of the two cuboids attached to $f$ and $f'$
(Figure~\ref{fillet}).  The faces of these cuboids attached to edges can
be classified into three types: two are adjacent to other such cuboids or
to the cuboids on $f$ and $f'$.  Two more are incident to the endpoints
of the shared edge and are again beveled to be nearly parallel to that
edge.  The final two point towards the interior of the polygon.  These
two faces are very close to parallel to each other, so that the two faces
incident to the endpoints of the shared edge have a ``kite''-like shape
resembling a slightly dented triangle.

Finally, we must cover the region near each vertex of $B$.  As seen from
the vertex, the faces of the cuboids we have already added form a vertex
figure that can be represented as an even polygon on the surface of a
small sphere centered at the vertex (Figure~\ref{kites}).  Each of the
sides of this polygon corresponds to one of the kite-shaped faces
incident to the vertex. We triangulate this even polygon, and add for
each interior diagonal of the triangulation another kite-shaped face, so
that the vertex neighborhood is partitioned into regions bounded by three
such faces. These regions correspond one-for-one
with the triangles in the triangulation. We then add to our buffer layer a
small cuboid in each such region.  Three faces
and seven vertices of the cuboid are already determined; the eighth
vertex is then fixed geometrically by the positions of the other seven. 
Since the three faces incident to the vertex of $B$ are all kite-shaped,
the three opposite faces are close to parallel to each other.  By making
all these cuboids attached to $B$ small enough, and by making their faces
close enough to parallel, this can all be done in such a way that no two
cuboids interfere with each other.

\item The second step of our topological method was to triangulate the
inner surface of the buffer layer, and
tetrahedralize the region inside this triangulated surface.
A tetrahedralization with $O(n^2)$ complexity can be found
using a method of Bern~\cite{Ber-SCG-93}.
(The bound claimed in that paper is $O(n+r^2)$ where $r$ is the number
of reflex edges, however our first step creates $\Theta(n)$ reflex edges.
Perhaps it is possible to use the information that many of these
edges are very close to flat, to reduce the complexity to depend
only on the reflex edges of $B$.)

\item The third step of our topological method was to
split each interior tetrahedron into four hexahedra.
In order to do this geometrically in a way consistent
across adjacent pairs of tetrahedra, we subdivide each
tetrahedron using planes through each edge and opposite midpoint
(Figure~\ref{hexedtet}).
It is not hard to show that these four planes meet in
a common point (e.g. by affine transformation
from the regular tetrahedron).
The subdivision on each tetrahedron face is therefore
along lines through each vertex and opposite midpoint.

\item The next step of our topological method
was to find a bipartition of $B$, and subdivide the interior edges
incident to one of the two vertex classes of the bipartition.
This step remains unchanged except that each vertex in the given
class may be incident to many interior edges; all are subdivided.

\item At this point, the cells of the buffer layer fall into
several classes.  The cells coming from faces of $B$
are like those of our topological construction,
polyhedra with seven quadrilateral facets and four hexagon facets.
The cells coming from edges of $B$ have four quadrilaterals,
three hexagons, and an octagon.
The cells coming from vertices of $B$ on one side of the
bipartition have 18 quadrilaterals and three hexagons.
The cells coming from the other side of the bipartition
have 18 quadrilaterals and three octagons.
In any case, the hexagon and octagon sides need to be subdivided,
in such a way that all cells end up with an even number of sides.
We can use the same idea of matching here; in fact
the cells at each vertex can be matched independently,
leaving one larger matching connecting the cells on faces and edges.

\item Finally, each buffer cell needs to be meshed.
This can be done independently for each cell,
but it would require a case analysis (which we have not done)
to show that each possible type cell can be meshed.
\end{enumerate}

Thus of the steps in our topological mesh generation procedure,
it is only the final finite case analysis which we have been unable
to extend to the geometric problem.

\section{Generalizations}

The only important property we used of topological balls
(with quadrilateral faces)
is that their boundaries form bipartite graphs; but the same extends to
simply connected domains with cubically meshed surfaces in any
dimension, as can easily
be seen via homology theory.  (Hetyai~\cite{Het-DCG-95} has an alternate
proof of bipartiteness for shellable complexes.)
Thus there seems no conceptual obstacle to extending this technique
to higher dimensional meshing problems, although it again
requires a case analysis or other technique such as that of Thurston and
Mitchell to prove that the resulting buffer cells are meshable.

An alternate direction for generalization is to more topologically complicated
polyhedra in three dimensions.  Mitchell~\cite{Mit-WCG-95} describes
a generalization of his method which applies whenever the input polyhedron
forms a handlebody that can be cut along evenly-many-sided disks to
reduce its complexity.  (Clearly, such a simplification can be used
independently of the mesh generation method to be used.)
Our method can handle an alternate class of polyhedra,
such as knot complements or bodies with disconnected boundaries,
for which no simplifying disk cut exists.
The only step where we used the connectivity of the input boundary
was in the result that a planar graph with even faces is bipartite;
instead we can simply require that the input polyhedron
be bipartite with evenly many sides in each boundary component.
We can topologically mesh any such polyhedron;
alternately, if we could solve the same finite set of cases as before
we can geometrically mesh any such polyhedron.
(The geometric case needs an extension of Bern's surface-preserving
tetrahedralization to non-simply-connected polyhedra,
due to Chazelle and Shouraboura~\cite{ChaSho-SCG-94}.)

\section{Conclusions}

We have shown that each simple polyhedron with evenly many quadrilateral
faces has a topological hexahedralization, and made some progress
towards finding geometric hexahedralizations.  Many questions remain
open.

\begin{itemize}
\item Can we find geometric hexahedralizations for all the cases arising
in our geometric hexahedralization technique?

\item If so, the result would be a method which generates $O(n^2)$
hexahedra.  Can this be reduced to $O(n+r^2)$ as has been done for
tetrahedralization~\cite{Ber-SCG-93, ChaPal-DCG-90, ChaSho-SCG-94}?

\item Which non-simple polyhedra admit topological hexahedralizations?

\item What is the worst case complexity of the Thurston-Mitchell
topological hexahedralization algorithm?

\item Is there any polyhedron which can be hexahedralized topologically
but not geometrically?

\item Can we make quality guarantees for hexahedral meshes similar to
those in our recent work on quadrilateral meshes \cite{BerEpp-IMR-97}?

\end{itemize}

\bibliography{hexmesh}
\end{document}